\documentclass[aps,preprint]{revtex4}%
\usepackage{amsfonts}
\usepackage{amsmath}
\usepackage{amssymb}
\usepackage{graphicx}%
\begin{document}
\title{Adiabatic demagnetization and generation of entanglement in spin systems}
\author{Gregory B. Furman, Victor M. Meerovich, and Vladimir L. Sokolovsky}
\affiliation{Physics Department, Ben Gurion University of the Negev, Beer Sheva, 84105
Israel Email:gregoryf@bgu.ac.il Phone:+97286472458 FAX:+97286472903}
\keywords{adiabatic demagnetization, entanglement }
\pacs{03.67.Mn, 76.60. -k}

\begin{abstract}
We study the entanglement emergence in a dipolar-coupled nuclear spin-1/2
system cooled using the adiabatic demagnetization technique. The unexpected
behavior of entanglement for the next- and next-next-neighbor spins is
revealed: entangled states of a spin system appear in two distinct temperature
and magnetic field regions separated by a zero-concurrence area. The magnetic
field dependence of the concurrence can have two maximums which positions are
determined by the initial conditions and number of spins in a chain.

\end{abstract}
\maketitle

\section{Introduction}

Prospects for practical application of quantum technology and devices based on
entangled state stimulate intensive qualitative and quantitative research of
quantum entanglement in various physical systems. Among these systems, a
system of nuclear spins presents excellent theoretical and experimental models
for studying the entanglement properties. Systems with a large number of spins
with various models of the spin interaction have been described in a number of
papers, see e.g. \cite{Amico2008,Horodecki2009,Osenda2005,S. I.
Doronin2007,Furman2009}, and the references therein. However, a major part of
the performed studies considers the temperature and field dependences without
taking into consideration the realistic cooling technique for achieving low
temperature at which entanglement appears. Adiabatic demagnetization (AD)
performed by variations of an external magnetic field is a very useful and
effective technique for the attainment of proper low temperatures, of the
order of microkelvins, at which entanglement appears in spin systems
\cite{C.J.Gorter1947,N. F. Ramsey1951,A. Abragam1957,A. Abragam1958,A.
Anderson1959,A. Abragam 1961}.

Recently, using adiabatic demagnetization in the rotating frame (ADRF)
\cite{C.P. Slichter1961,Goldman M 1970,Abragam1982} , entanglement in a
dipolar-coupled nuclear spin system was studied \cite{S I Doronin2009}. In
contrast to the conventional AD method, ADRF uses an external magnetic field
which much stronger than internal dipolar field and a radiofrequency field
with an offset from the Larmor frequency.

It was shown that there is no entanglement between remote spins, although the
considered model included dipole interactions between these spins which is the
typical condition to create entangled states.

In this paper we consider the entanglement emergence in a one-dimensional
dipolar-coupled nuclear spin-$1/2$ system cooled using the conventional AD
technique. We consider the case where the Zeeman energy is of the order of or
even less than the dipolar interaction one.

The structure of the paper is as follows: in the next section, we describe the
Hamiltonian for a spin system in an external field, conditions of the AD
realization and analyze variations of the spin temperature $T$ and
magnetization $M$ at AD. Then we consider the pairwise entanglement.
Discussion of the results is given in the final section. The numerical
calculation of the spin temperature, magnetization, heat capacity and
concurrence $C_{mn}$ between the $m$-th and $n$-th spins at arbitrary
orientation of the magnetic field is performed using the software based on the
MatLab package which allows us to consider systems of up to ten spins.

\subsection{Spin system at adiabatic demagnetization}

We consider a linear system of $N$ dipole-coupling nuclear spins $I=1/2$ in an
external magnetic field when the Zeeman energy of the order of or even less
than the dipolar interaction one. The Hamiltonian of the system can be
presented in the following form%
\begin{equation}
H=H_{z}+H_{dd} \tag{1}%
\end{equation}
where the Hamiltonian $H_{z}$ describes the Zeeman interaction between the
nuclear spins and external magnetic field%

\begin{equation}
H_{z}=\omega_{0}\sum_{k=1}^{N}I_{k}^{z}, \tag{2}%
\end{equation}
$\omega_{0}=\gamma\left\vert \vec{H}_{0}\right\vert $ is the energy difference
between the excited and ground states of an isolated spin, $\vec{H}_{0}$ is
the external magnetic field, $\gamma$ is the gyromagnetic ratio of a spin,
$I_{k}^{z}$ is the projection of the angular spin momentum operator on the
$z$- axes. The Hamiltonian $H_{dd}$ describing dipolar interactions in an
external magnetic field \cite{Abragam1982}:%

\begin{align}
H_{dd}  &  =\sum_{j<k}\frac{\gamma^{2}}{r_{jk}^{3}}\{\left(  1-3\cos^{2}%
\theta_{jk}\right)  \left[  I_{j}^{z}I_{k}^{z}-\frac{1}{4}\left(  I_{j}%
^{+}I_{k}^{-}+I_{j}^{-}I_{k}^{+}\right)  \right]  -\nonumber\\
&  \frac{3}{4}\sin2\theta_{jk}\left[  e^{-i\varphi_{jk}}\left(  I_{j}^{z}%
I_{k}^{+}+I_{j}^{+}I_{k}^{z}\right)  +e^{i\varphi_{jk}}\left(  I_{j}^{z}%
I_{k}^{-}+I_{j}^{-}I_{k}^{z}\right)  \right]  -\frac{3}{4}\sin^{2}\theta
_{jk}\left[  e^{-2i\varphi_{jk}}I_{j}^{+}I_{k}^{+}+e^{2i\varphi_{jk}}I_{j}%
^{-}I_{k}^{-}\right]  \} \tag{3}%
\end{align}
where $r_{jk}$, $\theta_{jk}$, and $\varphi_{jk}$ are the spherical
coordinates of the vector $\vec{r}_{jk}$ connecting the $j-$th and $k-$th
nuclei in a coordinate system with the $z$-axis along the external magnetic
field, $\vec{H}_{0}$, $I_{j}^{+}$and $I_{j}^{-}$ are the raising and lowering
spin angular momentum operators of the $j$-th spin. We consider the case when
\ $\omega_{0}\sim$ $D_{12}=\frac{\gamma^{2}}{r_{12}^{3}}$ (here $D_{12}$ is
the dipolar coupling constant for the nearest spins), and it is necessary to
take into account all the terms of the Hamiltonian, and not truncate any ones.

Let us analyze conditions of the adiabatic demagnetization in the spin
system.To prevent heat exchange between a spin system and a lattice, the
evolution time of the considered spin system has to be much shorter than the
spin-lattice relaxation time, $T_{1}$. On the other hand, decoherence in a
spin system refers to how the system loses the quantum coherence features and
can be characterized by time of the order of $T_{2}$, the lifetime of the free
precession signal ($T_{2}<<T_{1}$ \cite{A. Abragam 1961,C.P. Slichter1961}).
Decoherence times, $T_{2}\sim\left(  \gamma H_{loc}\right)  ^{-1}$( $H_{loc}$
\ is the local magnetic field created by spins) for spin systems at low
temperature typically range between nanoseconds and seconds \cite{David P.
DiVincenzo (1995)}. In contrast to $T_{2}$, $T_{1}$ takes values {}{}in the
range from minutes to hours \cite{A. Abragam 1961,C.P. Slichter1961}. For
example, it was found in LiF at $T=2$K that $T_{1}=50$ min for $^{19}F$ and
$T_{1}=$15 hours for Li \cite{A. Abragam1958}. If to choose the characteristic
time of variation of the external field such that $t_{ch}>$ $T_{2}$ , the
system can be considered as being in thermal equilibrium at every point of
time and its density matrix is%

\begin{equation}
\rho=\frac{1}{Z}\exp\left\{  -\frac{\beta H}{D_{12}}\right\}  . \tag{4}%
\end{equation}
where $\beta$ is the inverse spin temperature in units of $D_{12}$,
$\beta=\frac{D_{12}}{k_{B}T}$, $T$ is the spin temperature and $Z=Tr\left\{
\exp\left(  -\frac{\beta H}{D_{12}}\right)  \right\}  $ is the partition
function that encodes the statistical properties of a system in thermodynamic equilibrium.

A spin system can be considered in the thermal equilibrium during evolution if
\cite{A. Abragam1958}%

\begin{equation}
\frac{dH_{0}}{dt}<<\gamma H_{loc}^{2}. \tag{5}%
\end{equation}
The mean local magnetic field $H_{loc}$ determines by the following expression
$H_{loc}=\frac{1}{\gamma}\sqrt{\frac{Tr\left(  H_{dd}^{2}\right)  }%
{TrI_{z}^{2}}}$ and depends on the number of spins in the system \cite{A.
Abragam1958}. Our calculations of this field for all cases considered below,
$N=2\div10$, give $H_{loc}=\left(  0.8\div1.2\right)  D_{12}$, where
$D_{12}\sim10$ G \cite{A. Abragam1958}, which leads to the value of the order
of few gausses. For example, in LiF calculation gives $H_{loc}=7.77$ G
\cite{A. Abragam1958}. With this value and $\gamma=2\pi\times4005.5$
$\frac{Hz}{G}$\ Eq (5) gives that $\frac{dH_{0}}{dt}<<$ $1.5\times10^{6}%
\frac{G}{s}$ . \ If \ we take $\frac{dH_{0}}{dt}=1.5\times10^{5}$ $\frac{G}%
{s}$ the field change time from $6000$ G to $10$ G equels $t_{ch}$
$=4\times10^{-2}$ s that satisfies the condition of adiabaticity
$T_{1}>>\ t_{ch}\geq\ T_{2}\sim10^{-5}$ s.

Thermodynamic variables of the system, such as entropy, $S$ and heat capacity
$C$, can be expressed in terms of the partition function or its derivatives.
For example, the entropy $S$ and heat capacity $C$ are given by
\begin{equation}
S=k_{B}\frac{\partial\left(  \beta\ln Z\right)  }{\partial\beta} \tag{6}%
\end{equation}
and
\begin{equation}
C=k_{B}\beta^{2}\frac{\partial^{2}\ln Z}{\partial\beta^{2}}, \tag{7}%
\end{equation}
respectively. The magnetization $M$ of the spin system can be also expressed
in the terms of the partition function%

\begin{equation}
M=-\frac{1}{\beta Z}\frac{\partial Z}{\partial B} \tag{8}%
\end{equation}
where $B$ is the external magnetic flux density.

The variation of the parameters at AD is simulated by the following way. At an
initial condition $\beta=\beta_{in}$ and $H_{0}=H_{0in}$ the entropy is
calculated using Eqs. (4) and (6), the external magnetic field is decreased
and a new temperature is determined from the condition $S=const$.

Fig. 1 shows the inverse spin temperature versus the external magnetic field.
One can see that the inverse spin temperature $\beta$ is proportional to
$\omega_{0}$ at $\omega_{0}>>D_{12}$\ for all the spin systems considered
here. For $\omega_{0}<D_{12}$the inverse spin temperature is practically
independent of $\omega_{0}$ (see the inset of Fig.1). The main change of
temperature occurs in the range of \ $\omega_{0}/D_{12}$ from $5$ till $1$. At
$\omega_{0}/D_{12}>5$ the inverse spin temperature increases approximately
linearly with a decrease of the magnetic field. In contrast to the dependence
of the temperature on the external magnetic field, the magnetization is about
constant at $\omega_{0}>5D_{12}$and decreases linearly with the magnetic field
decrease at $\omega_{0}<D_{12}$ (see the inset of Fig.2). At high fields
$\omega_{0}>3D_{12}$, the temperature and magnetization became independent of
the number of spins in the system.

In the region of high temperatures (low $\beta$), our results coincide with
well-known dependencies obtained in the framework of the high temperature
approximation $\left\vert \frac{\beta H}{D_{12}}\right\vert <<1$ for $N>>1$
(see dark yellow shot dash-dotted lines in Figs. 1 and 2) \cite{A.
Abragam1958,C.P. Slichter1961}. We can see from Fig. 2 that the high
temperature approximation is valid for $\omega_{0}$ $>D_{12}$ which
corresponds to \ $\beta<2.3$ (Fig.1). Below we will consider entanglement
measures and temperature and field dependence of entanglement of the linear
chain of dipolar coupled spins under AD.

\subsection{Generation of entanglement at adiabatic demagnetization}

There are several parameters characterizing the entangled state of a spin
system: the von Neumann entropy, entanglement of formation, log negativity,
concurrence of a pair of spins, and etc. \cite{Amico2008,Horodecki2009,W. K.
Wootters1998,G. Vidal2000,G. Vidal2002,Berry2006}. We will characterize the
entangled states by the concurrence between two, $m$-th and $n$-th, spins
which is defined as \cite{W. K. Wootters1998}%

\begin{equation}
C_{mn}=\max\left\{  q_{mn},0\right\}  , \tag{9}%
\end{equation}
with $q_{mn}=\lambda_{mn}^{\left(  1\right)  }-\lambda_{mn}^{\left(  2\right)
}-\lambda_{mn}^{\left(  3\right)  }-\lambda_{mn}^{\left(  4\right)  }$. Here
$\lambda_{mn}^{\left(  k\right)  }\left(  k=1,2,3,4\right)  $ are the square
roots of eigenvalues, in descending order, of the following non-Hermitian
matrix:
\begin{equation}
R_{mn}=\rho_{mn}\left(  \sigma_{y}\otimes\sigma_{y}\right)  \tilde{\rho}%
_{mn}\left(  \sigma_{y}\otimes\sigma_{y}\right)  , \tag{10}%
\end{equation}
where $\rho_{mn}$ is the reduced density matrix. For the $m$-th and $n$-th
spins, the reduced density matrix $\rho_{mn}$ is defined as \ \ $\rho
_{mn}=Tr_{mn}\left(  \rho\right)  $ where $Tr_{mn}\left(  ...\right)  $
denotes the trace over the degrees of freedom for all spins except the $m$-th
and $n$-th spins. In Eq. (10) $\tilde{\rho}_{mn}$ is the complex conjugation
of the reduced density matrix \ $\rho_{mn}$ and $\sigma_{y}$ is the Pauli
matrix $\sigma_{y}=%
\begin{array}
[c]{cc}%
0 & -i\\
i & 0
\end{array}
$. For maximally entangled states, the concurrence is $C_{mn}=1$ while for
separable states $C_{mn}=0$. The concurrence for two-spin system in the
thermal equilibrium is zero at $\theta=0$ and $\theta=\pi$, and it reaches the
maximum values at $\theta=\frac{\pi}{2}$ and $\varphi=0$ \cite{Furman2011}. We
have confirmed these results by numerical calculations for all spin pairs in
the chains with $N=3,\ldots,10$. Therefore, we will restrict ourselves to the
cases corresponding to the concurrence maximum.

Fig. 3 presents the results of numerical calculations of the concurrence and
heat capacity as functions of the magnetic field at AD. With decreasing the
magnetic field, the both quantities increase up to the maximum and then
decrease till zero. Both, the concurrence and heat capacity decrease with
increasing number of spins for and are practically independent of the spin
number for .

Appearance of entanglement at AD and behavior of the concurrence depend on
initial conditions: initial values of the magnetic field and temperature from
which AD is started. This dependence is illustrated for a two-spin chain by
the phase diagram of Fig. 4. If AD starts at point $A$ (Fig.4) above the curve
separating the entangled and separable states, a non-zero concurrence is
observed till point $a_{1}$ where the AD phase trajectory crosses the boundary
line (Figs. 4 and 5). Starting from point $B$ we achieve the entangled state
between points $b_{1}$ and $b_{2}$, while starting from point $C$ does not
lead to appearance of the entangled state. The similar results are obtained
for all neighboring spins in the chains with $N=3,\ldots,10$.

The results for remote spins are presented in Figs. 6-8. At $\omega_{0}%
/D_{12}>5.5$, the concurrence for the next-neighbor spins $C_{13}$\ has two
maximums which are separated by area with zero concurrence (Figs. 6 and 7).
The zero concurrence area expands with increasing the spin number in a chain
(Fig. 6) and the initial temperature (Fig. 7b) and reduces when the initial
magnetic field increases (Fig. 7a). At AD the first maximum (at higher field)
decreases with the increase of the spin number in a chain (Fig. 6) and of the
initial magnetic field and temperature (Fig.7). The first maximum position
moves towards higher magnetic fields with increasing spin number and initial
temperature. The second maximum also moves towards higher magnetic fields with
increasing the initial field (Fig. 6) but its position does not practically
depend on the initial temperature (Fig.7).

The concurrence for next-next-neighbor spins $C_{14}$ also has two maximums
while only one maximum presents in the magnetic field dependence of the
concurrence for the first and fifth spins $C_{15}$ (Fig. 9). In all considered
cases, the concurrency for remote spins practically does not depend on the
number of spins for $N>7$.

\bigskip

\subsection{Discussion and conclusion}

We have investigated entanglement in a linear chain of dipole-coupling spins
$s=1/2$ at $AD$ which is one of the ways to achieve ultra-low spin temperatures.

The inverse temperature at which entanglement appears is $\beta\sim2.3$ . Let
us estimate this temperature for fluorine in calcium-fluoride CaF$_{2}$ with
the dipolar interaction energy of the order of a few kHz (in frequency units)
\cite{Abragam1982}. Taking as in \cite{Abragam1982,M. Goldman1974} $H_{0}=3$ G
we have $\omega_{0}=12$ kHz, which leads to $T=0.34$
$\mu$%
K. The estimated value of temperature is in good agreement with those reported
early for the spin system $s=1/2$ with the $XY$ Hamiltonian \cite{S. I.
Doronin2007} and for the Hamiltonian of two dipolar coupling spins
\cite{Furman2011}. It is interesting that the transition to the ordered
states, such as antiferromagnetic, of nuclear spins was observed in a single
crystal of calcium-fluoride precisely at $T=0.34$%
$\mu$%
K \cite{Abragam1982,M. Goldman1974}.

An unexpected behavior of the next nearest-neighbor and next-next
nearest-neighbor concurrences was obtained: remote spins in the thermal
equilibrium are entangled in two distinct temperature and magnetic field
regions (Figs 6-8). In most studied cases the ground state at zero temperature
is entangled \cite{Amico2008,Horodecki2009,Benenti2007}. However examples were
found \cite{Berry2006,M. A. Nielsen} where the ground state is separable and
there are two temperature regions of entangled states. We show that the same
behavior takes place also for the field dependence of the concurrence (Fig.6)
at certain initial values of the temperature and magnetic field (Fig. 4).

Our investigation demonstrates that the qualitative behavior of entanglement
with temperature and magnetic field can be much more complicated than might
otherwise have been expected. Thus, the AD technique allows one to generate
entangled states between dipolar coupling spins in a linear chain. It opens a
simple and effective way to the experimental testing of entanglement in spin systems.

\bigskip

Figure captions

\bigskip

Fig. 1 Fig. 1 Inverse spins temperature as a function of applied field, , at
AD in chains with various number of spins: black solid line - $N=4$, red
dashed line -$N=5$, green dotted line -$N=6$, blue dash-dotted line -$N=7$,
cyan dash-dot-dotted line -$N=8$, magenta shot-dashed line - $N=9$, yellow
shot-dotted - $N=10$, dark yellow shot-dash-dotted -- the high temperature
approximation for $N>>1$.

\bigskip

Fig. 2 Magnetization, as a function of applied field for an adiabatic
demagnetization in chain with various number of spins: black solid line -
$N=4$, red dashed line - $N=5$\ green dotted line -$N=6$, blue dash-dotted
line \ \ \ \ \ $N=7$\ \ \ \ \ \ cyan dash-dot-dotted line --$N=8$, magenta
shot-dashed line -- $N=9$, yellow shot-dotted $N=10$, dark yellow
shot-dash-dotted -- the high temperature approximation for $N>>1$.

\bigskip

Fig.3 Field dependence of concurrence between the first and second spins (a)
and of heat capacity C per spin (b) in various chains: black solid line
-$N=2$, red dashed line - $N=3$, green dotted line - $N=4$, blue dot-dashed
line - $N=5$, cyan dash-dot-dotted line - $N=6$, magenta shot dashed line -
$N=7$, yellow shot dotted - $N=8$, dark yellow shot dash-dotted - $N=9$, navy
- $N=10$.

\bigskip

Fig. 4 The phase diagram for a two-spin system at AD started from various
initial points: $A$\{ $\omega_{0}=2.4,\beta=1.06$\}; $B$\{ $\omega_{0}=3$,
$\beta=0.7$\}; C\{$\omega_{0}=2.7$, $\beta=0.443$\}. The black solid line is
the boundary between the entangled and separable states; green dashed, red
dotted, and blue dot-dashed lines show the phase trajectories started from
$A$, $B$, $C$ respectively.

Fig. 5 Field dependence of the concurrence at AD starting from initial
conditions corresponding to $A$\{$\omega_{0}=2.4$, $\beta=$1.06\} (green
dashed line) and $B$\{ $\omega_{0}=3$, $\beta=0.7$\} (red dotted line) in the
$\omega_{0}-\beta$ diagram of Fig. 4.

\bigskip

Fig.6 Field dependence of concurrence for the first and third spins in various
chains at $AD$ starting from $\beta$ $=2$ : black solid line - $N=3$, red
dashed line -$N=4$ green dotted line -$N=5$, blue dash-dotted line -$N=6$ cyan
dash-dot-dotted line --$N=7$, magenta shot-dashed line -- $N=8$, yellow
shot-dotted $N=9$, dark yellow shot-dash-dotted -- $N=10$.

\bigskip

Fig.7 Field dependence of concurrence for the first and third $C_{13}$ spins
in 3-spin chain (a) and 6-spin chain (b) at AD starting from various initial points.

a) - initial inverse temperature $\beta$ $=1$ and various initial external
magnetic fields: black solid line -- $\omega_{0}/D_{12}=5$; red dashed line -
$\omega_{0}/D_{12}=5.3$; green dotted line - $\omega_{0}/D_{12}=5.5$; blue
dash-dotted line - $\omega_{0}/D_{12}=6$.

b)- initial external magnetic field $\omega_{0}/D_{12}=20$ and various initial
inverse temperatures: black solid line - $\beta=0.3$; red dashed line -
$\beta$ $=0.4$; green dotted line - $\beta$ $=0.6$; blue dash-dotted line -
$\beta$ $=0.9$; cyan dash-dot-doted line -- $\beta$ $=1$; magenta shot-dashed
line -- $\beta$ $=3$.

Fig. 8 Field dependence of concurrence $C_{14}$ for the first and fourth spins
in various chains: (a) $N=4$; (b) black solid line - $N=5$, red dashed line -
$N=6$ green dotted line -$N=7$ for the initial inverse temperature
$\ \ \beta=2$ and magnetic field $\omega_{0}/D_{12}=40$.

\bigskip

Fig. 9 Field dependence of concurrence $C_{15}$ for the first and fifth spins
in various chains: black solid line - $N=5$, red dashed line -$N=6$, green
dotted line -$N=7$; for the initial inverse temperature \ \ \ \ \ \ \ $\beta
=2$\ \ \ \ \ \ \ \ \ \ and magnetic field $\omega_{0}/D_{12}=40$.

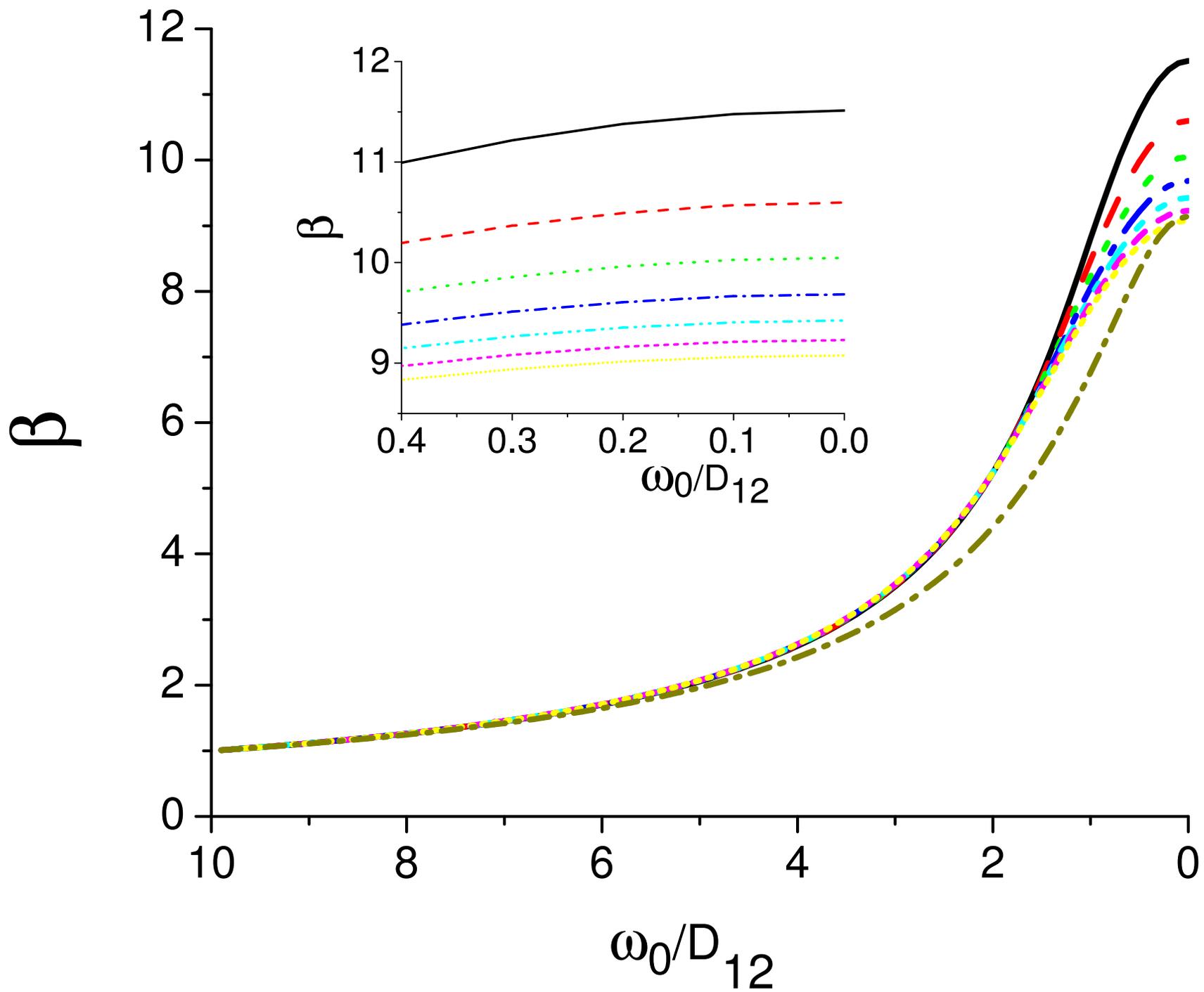
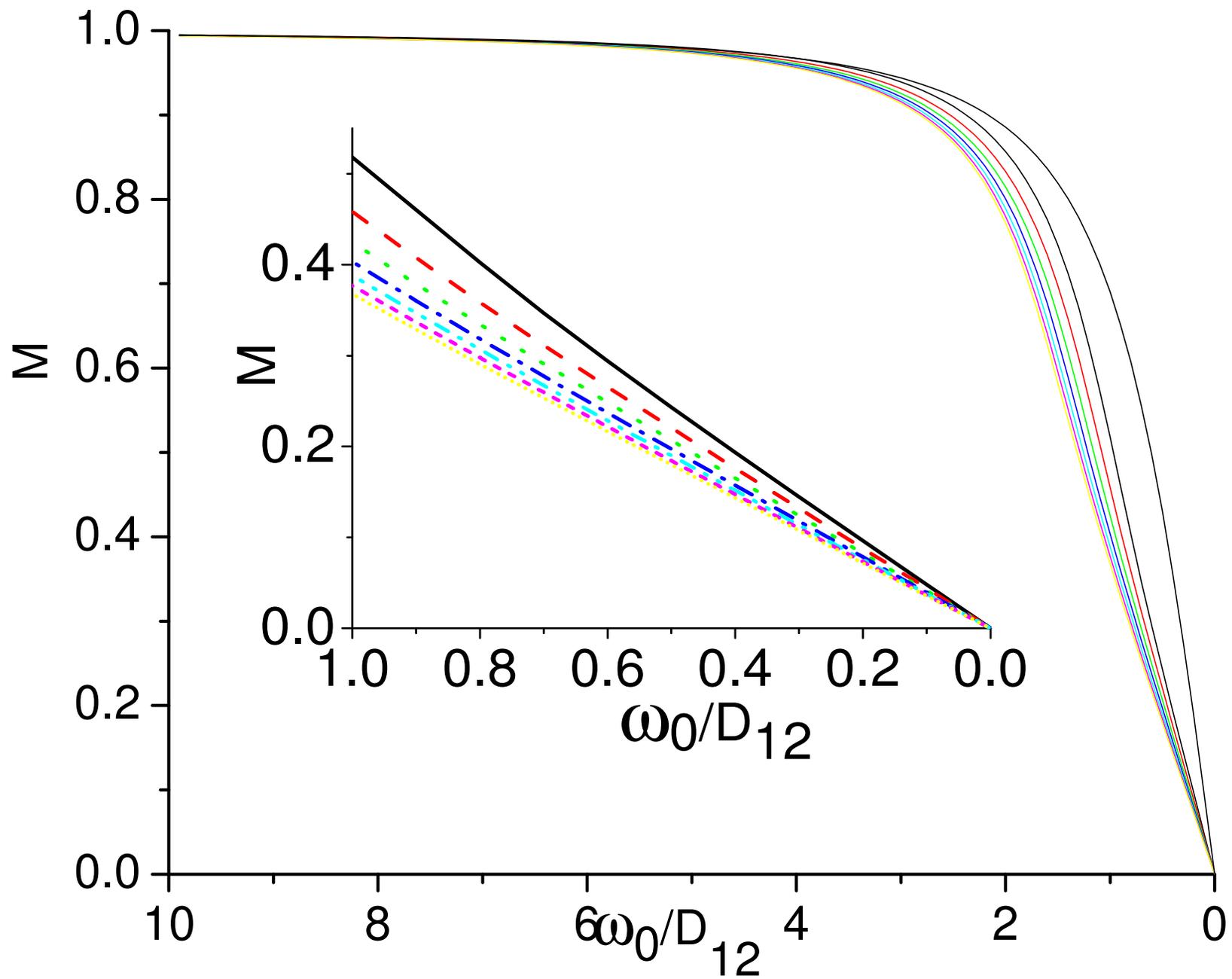
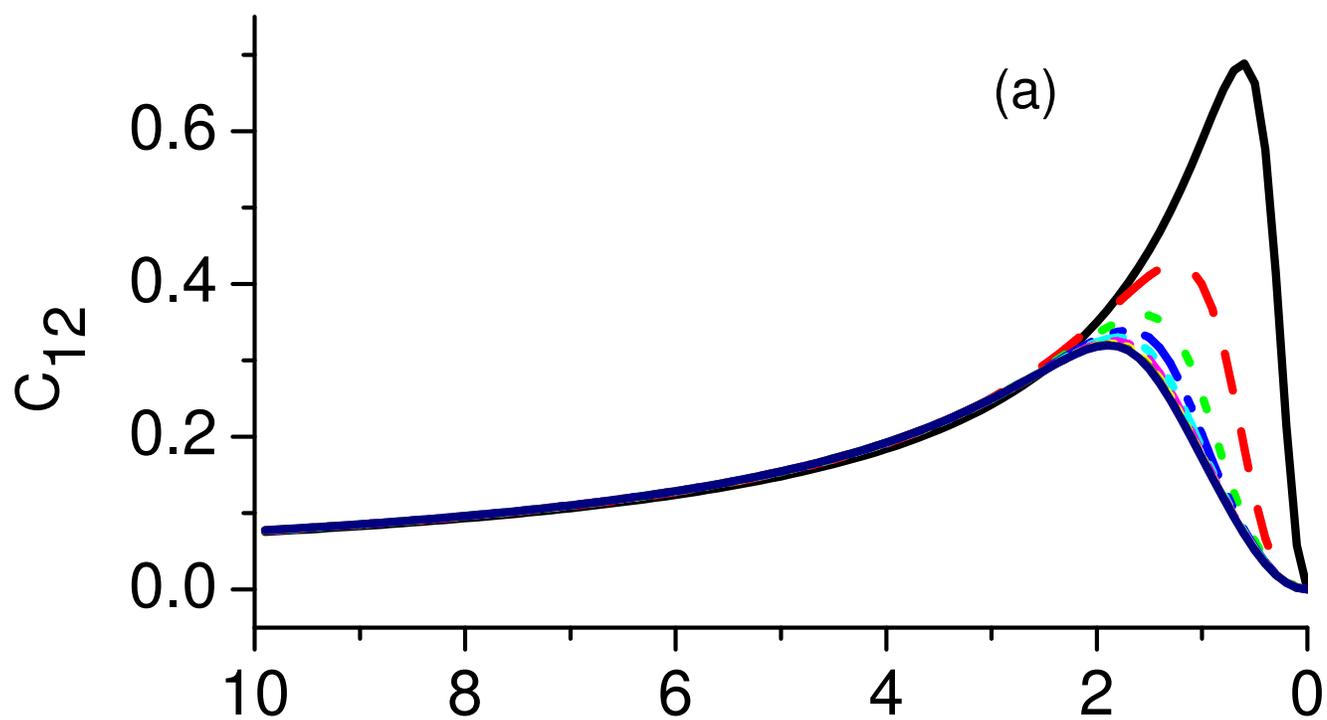
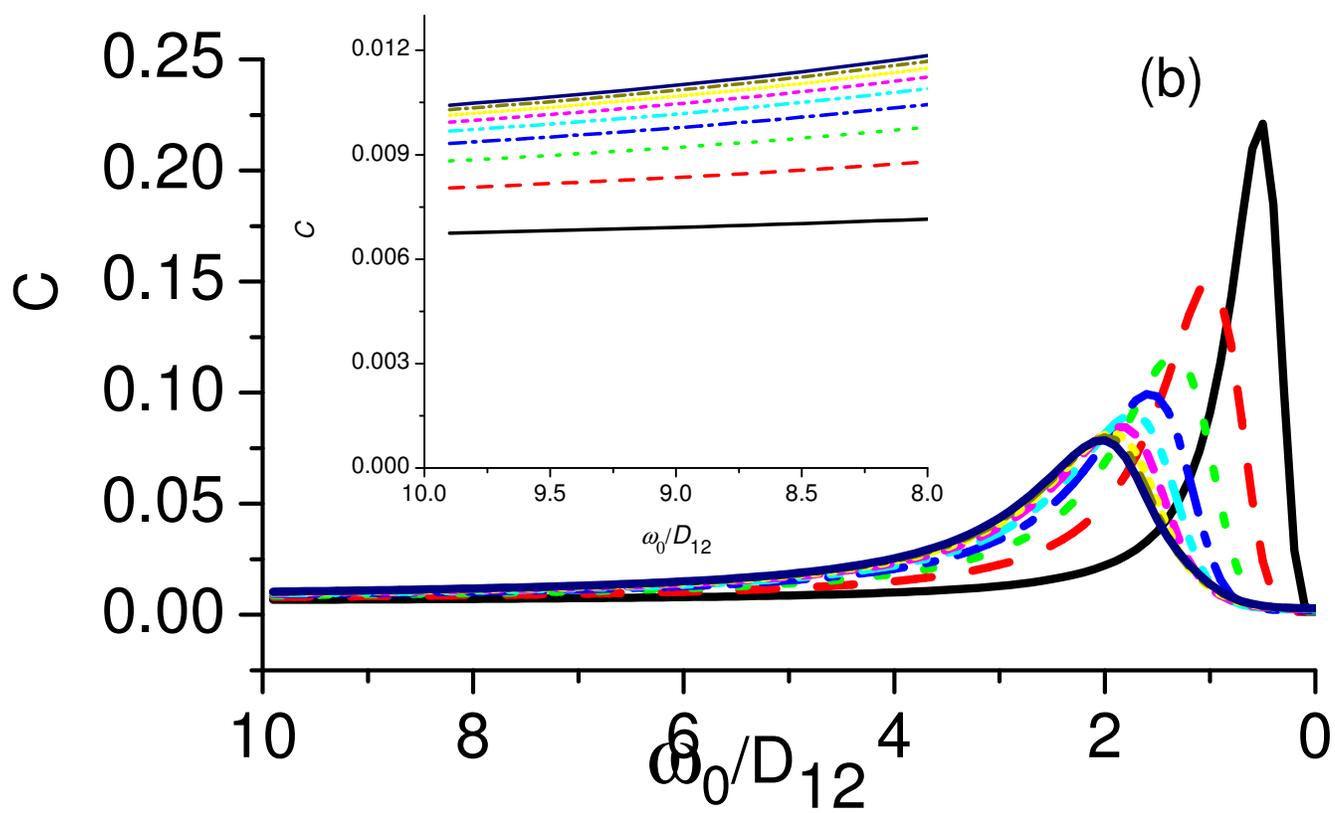
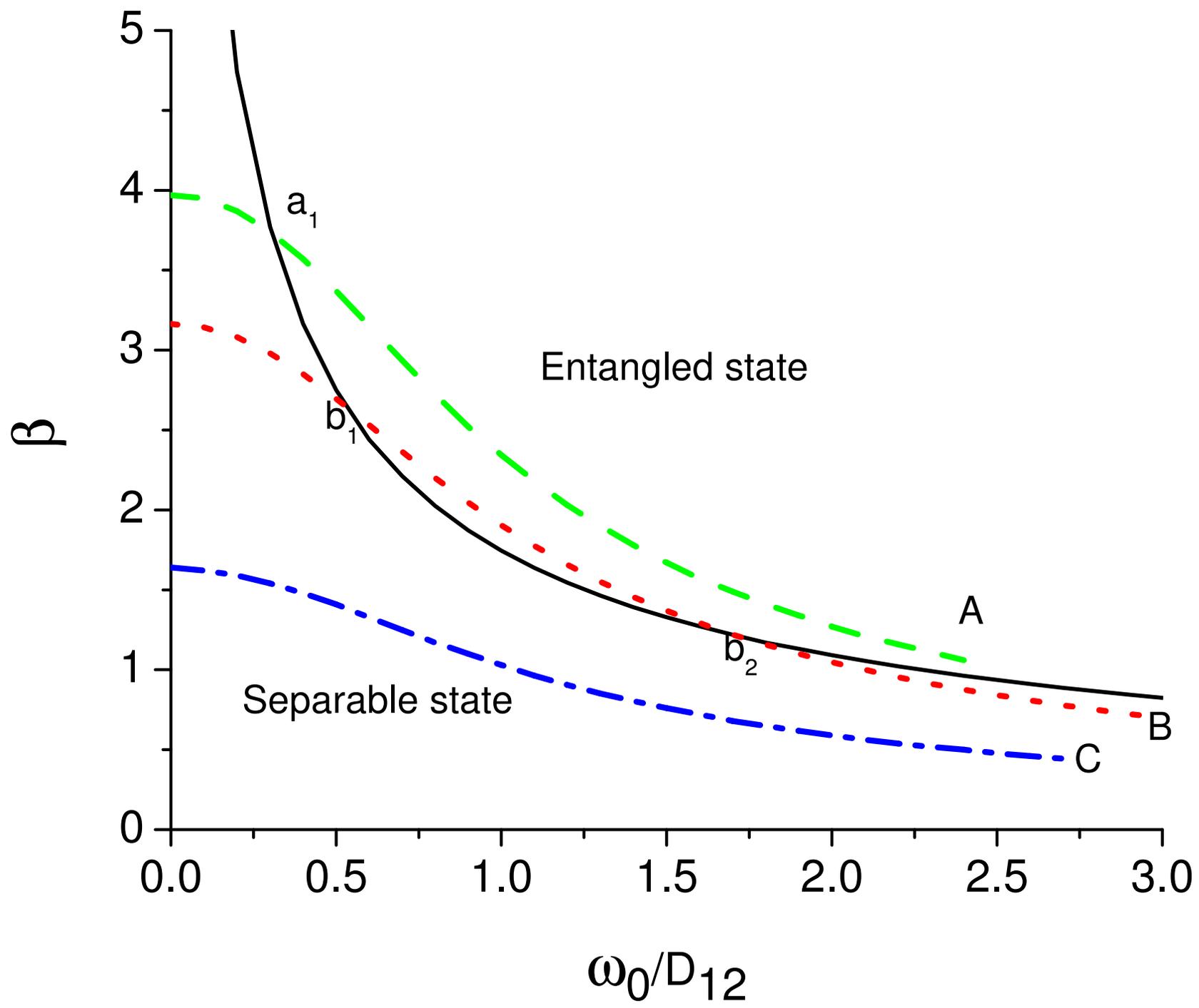
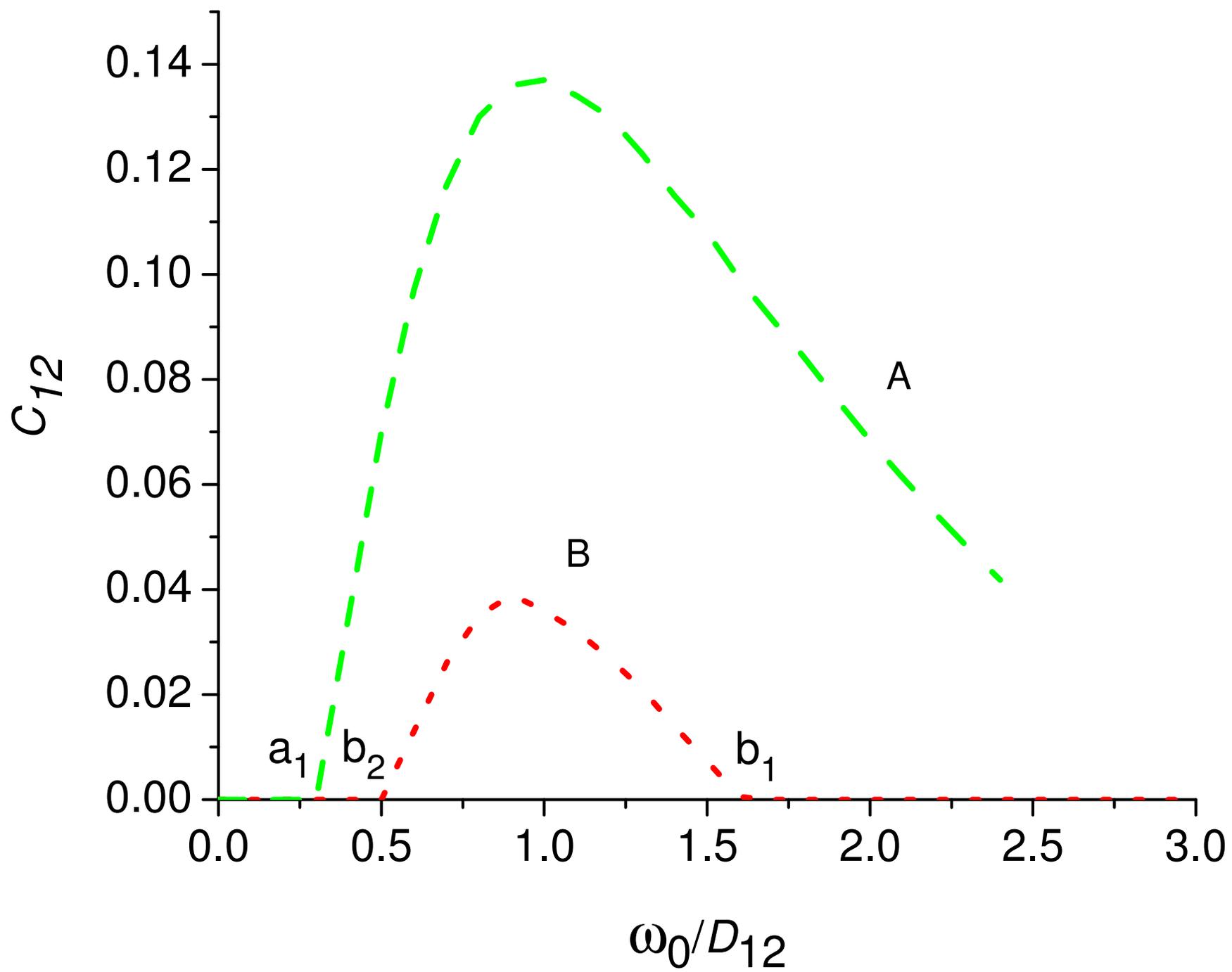
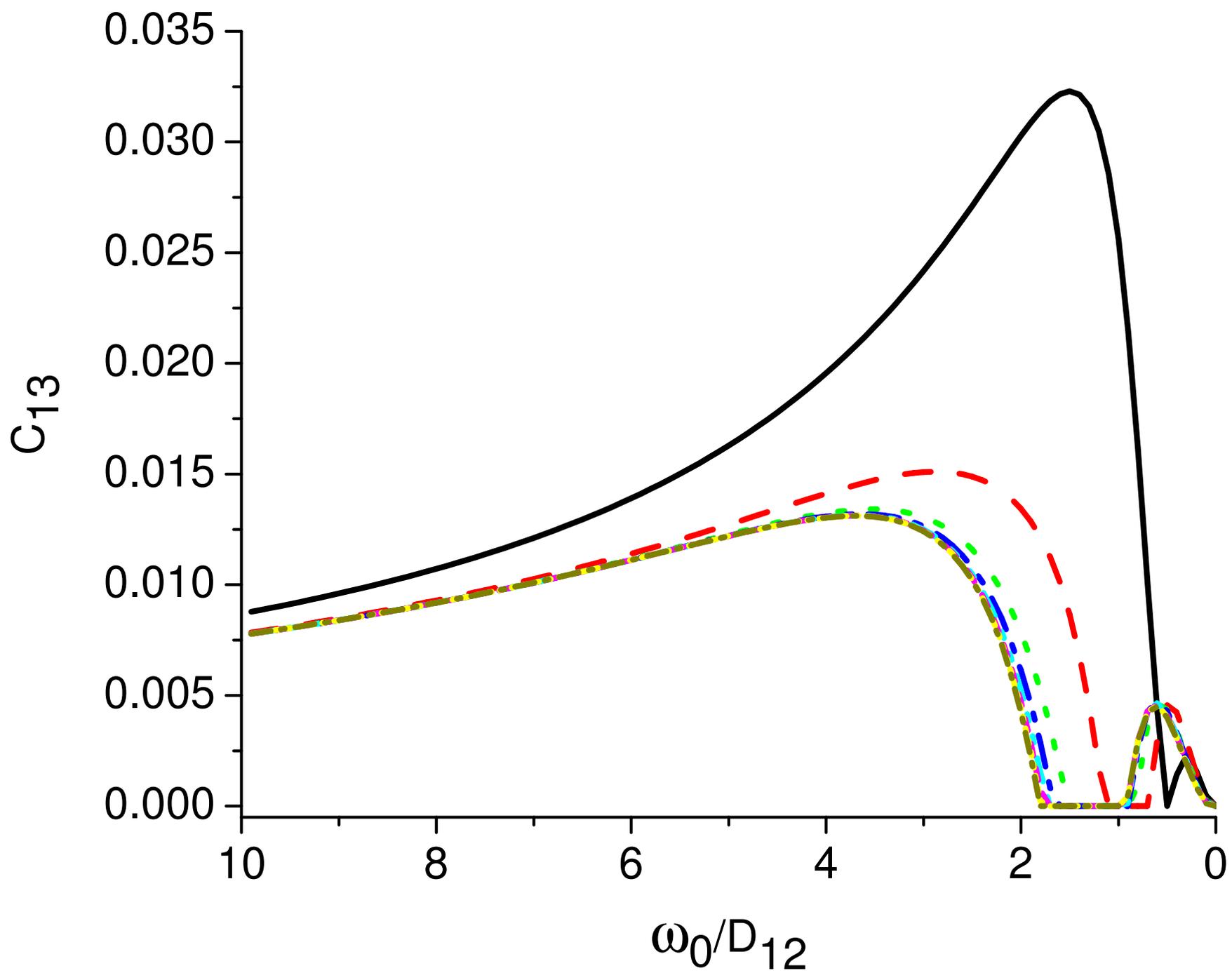
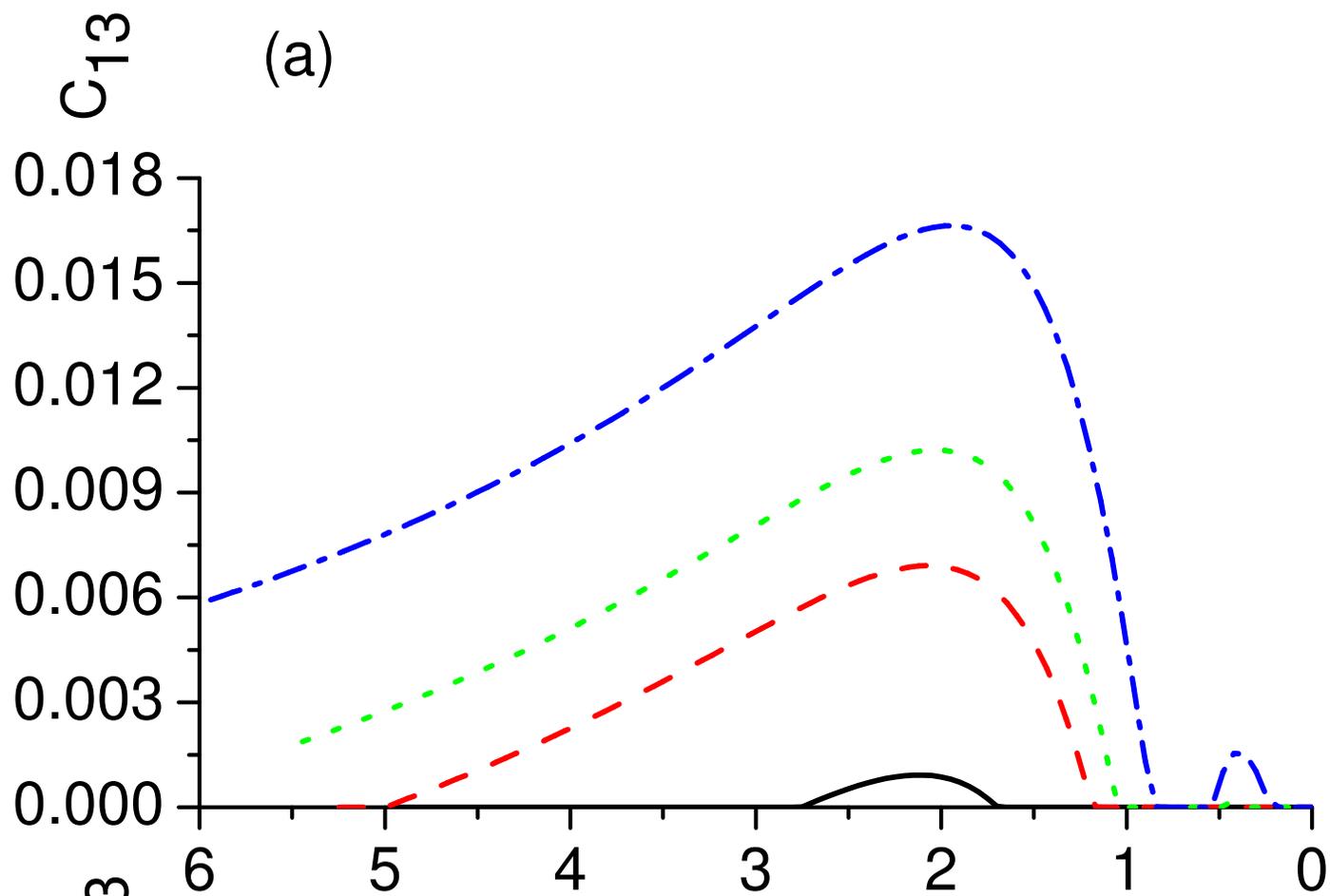
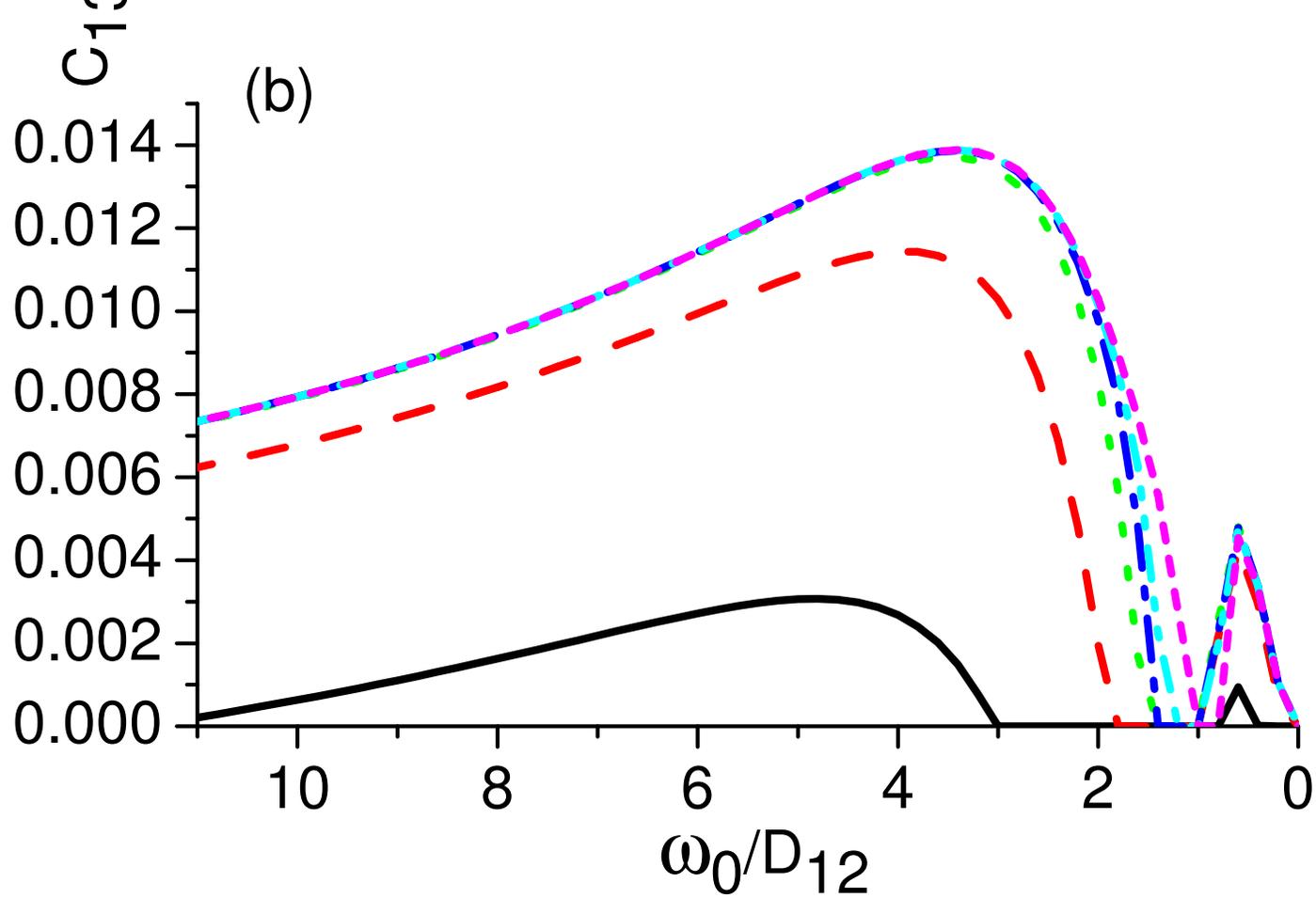
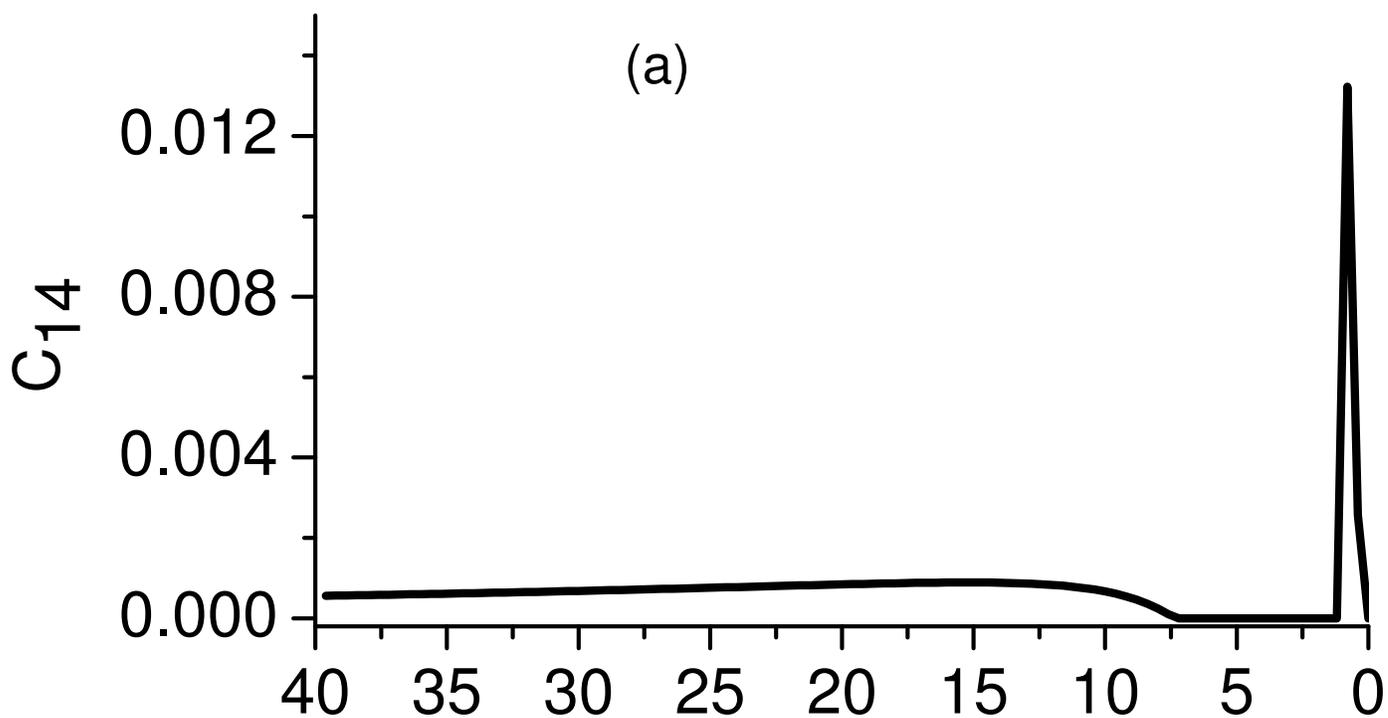
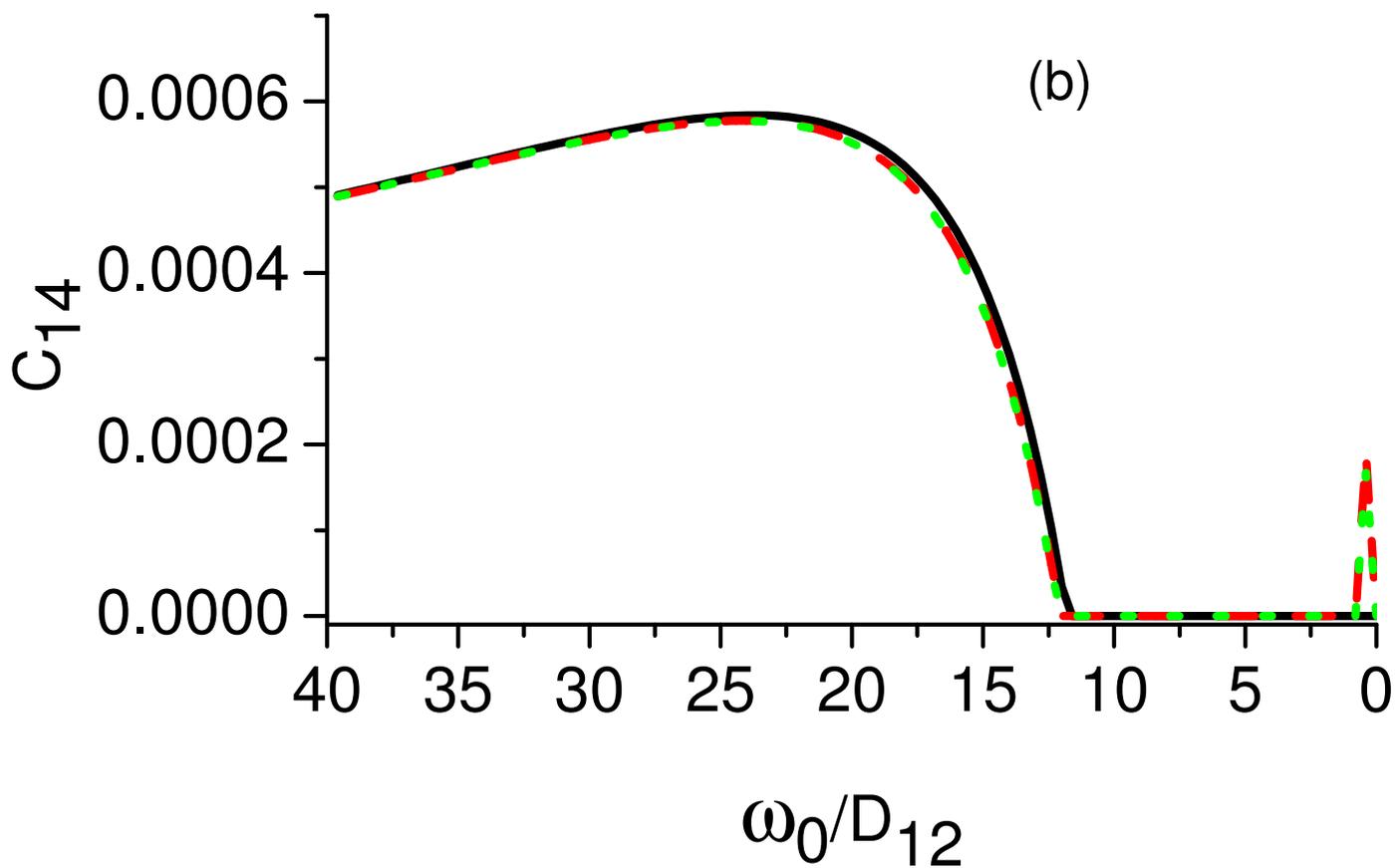
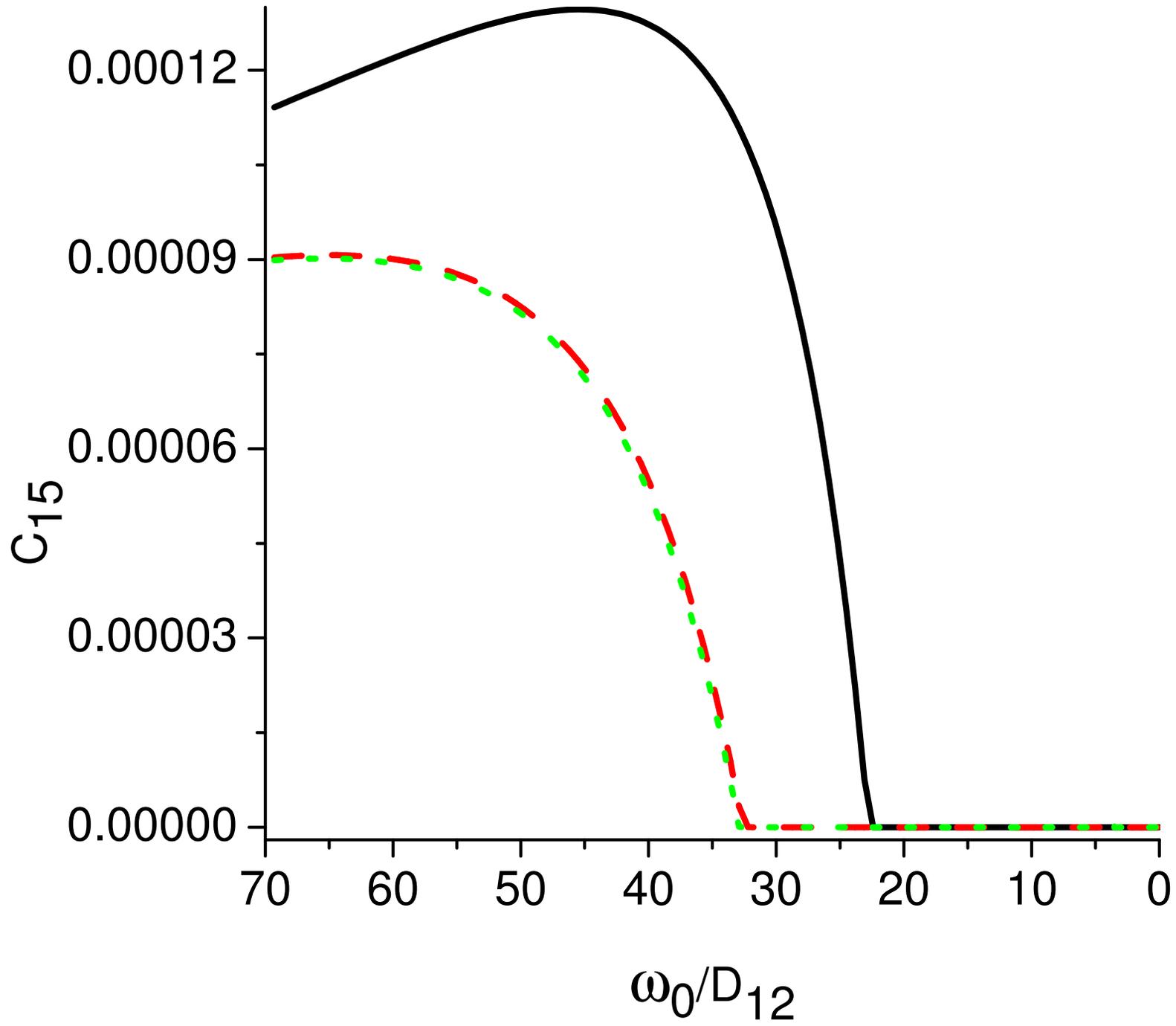
\end{document}